\documentclass[seceq]{ptptex}
\usepackage[dvips]{graphicx,color}
\usepackage{multirow}
\usepackage{bm}
\usepackage{wrapft}

\def\dfrac#1#2{{\displaystyle\frac{#1}{#2}}}

\def\beq{\begin{equation}}
\def\eeq{\end{equation}}
\def\bea{\begin{eqnarray}}
\def\eea{\end{eqnarray}}

\def\ga{\mathrel{\mathpalette\fun >}}
\def\fun#1#2{\lower3.6pt\vbox{\baselineskip0pt\lineskip.9pt
\ialign{$\mathsurround=0pt#1\hfil##\hfil$\crcr#2\crcr\sim\crcr}}}
\title{%        %You can use \\ for explicit line-break.
Three-Body Model Analysis of Subbarrier $\alpha$ Transfer Reaction
}

\author{%       %Use \scshape for the family name.
Tokuro \textsc{Fukui},$^{1}$\thanks{E-mail: fukui@phys.kyushu-u.ac.jp}
Kazuyuki \textsc{Ogata},$^{1}$\thanks{Present address: Research Center for Nuclear Physics, Osaka University}
and
Masanobu \textsc{Yahiro}$^{1}$
}

\inst{%     %Affiliation, neglected when [addenda] or [errata].
$^{1}$Department of Physics, Kyushu University,
Fukuoka 812-8581, Japan
}

\abst{%         %This abstract is neglected when [addenda] or [errata].
Subbarrier $\alpha$ transfer reaction
$^{13}$C($^{6}$Li$,d$)$^{17}$O(6.356~MeV, $1/2^+$) at 3.6 MeV
is analyzed with a $\alpha+d+^{13}$C three-body model,
and the asymptotic normalization coefficient
(ANC) for $\alpha + ^{13}$C $\longrightarrow$ $^{17}$O(6.356~MeV, $1/2^+$),
which essentially determines the reaction rate of
$^{13}$C($\alpha,n$)$^{16}$O, is extracted.
Breakup effects of $^6$Li in the initial channel and those of
$^{17}$O in the
final channel are investigated with the continuum-discretized
coupled-channels method (CDCC). The former is found to have a large
back-coupling to the elastic channel, while the latter turns out
significantly small. The transfer cross section calculated with Born
approximation to the transition operator, including breakup
states of $^6$Li, gives
$(C_{\alpha ^{13}\mbox{\scriptsize C}}^{^{17}\mbox{\scriptsize O}^*})^2
=1.03 \pm 0.29$~fm$^{-1}$.
This result is consistent with the value obtained by the previous DWBA
calculation.
}

\begin{document}
\maketitle

\section{Introduction}
\label{sec1}

Transfer reactions below Coulomb barrier energies
are known to be a powerful
technique to determine asymptotic properties of the overlap
between the
initial and final state wave functions, essentially free from
uncertainties associated with optical potentials and structural
complexity of wave functions in the nuclear interior
region.~\cite{Satchler} Recently, subbarrier $\alpha$ transfer
reactions have been used to
indirectly measure cross sections of $\alpha$-induced
reactions of astrophysical interest.~\cite{Eric1,Eric2}
In Ref.~\citen{Eric1},
Johnson and collaborators determined the reaction rate of
$^{13}$C($\alpha,n$)$^{16}$O by measuring the
$^{13}$C($^{6}$Li$,d$)$^{17}$O(6.356~MeV, $1/2^+$) reaction;
for simplicity, we henceforth denote the final state of $^{17}$O
as $^{17}$O$^*$.
The $^{13}$C($\alpha,n$)$^{16}$O reaction is considered to be
important as a neutron source for the slow neutron capture process
(s-process) taken place in the asymptotic giant branch (AGB)
stars.~\cite{Iben}

In the cross section formula, Eq.~(1) of Ref.~\citen{Eric1},
of the $^{13}$C($\alpha,n$)$^{16}$O reaction
based on $R$-matrix approach~\cite{TM}, the asymptotic normalization
coefficient (ANC) for
$\alpha + ^{13}$C $\longrightarrow$ $^{17}$O$^*$,
$C_{\alpha ^{13}\mbox{\scriptsize C}}^{^{17}\mbox{\scriptsize O}^*}$,
is the only missing quantity. Throughout this study we consider
the ANC with Coulomb-modification,~\cite{Eric1} i.e., a value divided by
the Gamma function $\Gamma (2+\eta)$, where $\eta$ is
the Sommerfeld parameter for the $\alpha$-$^{13}$C system.
In Ref.~\citen{Eric1}, the $\alpha$ transfer reaction
$^{13}$C($^{6}$Li$,d$)$^{17}$O$^*$
was analyzed with DWBA, disregarding the breakup effects of
$^6$Li and $^{17}$O$^*$, and
$(C_{\alpha ^{13}\mbox{\scriptsize C}}^{^{17}\mbox{\scriptsize O}^*})^2
=0.89 \pm 0.23$~fm$^{-1}$
was extracted.
The ground state energy of $^{6}$Li is, however, just 1.47 MeV below
the $\alpha+d$ threshold. Furthermore, the
binding energy of $^{17}$O$^*$, i.e.,
$^{17}$O(6.356~MeV, $1/2^+$), from the
$\alpha+^{13}$C threshold is only 3~keV. Therefore, to extract
a reliable value of
$C_{\alpha ^{13}\mbox{\scriptsize C}}^{^{17}\mbox{\scriptsize O}^*}$,
one should investigate how important $^6$Li and $^{17}$O$^*$ breakup are
in the $\alpha$ transfer reaction.

The purpose of the present Letter is to analyze the
$^{13}$C($^{6}$Li$,d$)$^{17}$O$^*$ reaction at
3.6 MeV (for the incident energy of $^{6}$Li)
with the three-body ($\alpha+d+^{13}$C) model
and to determine
$C_{\alpha ^{13}\mbox{\scriptsize C}}^{^{17}\mbox{\scriptsize O}^*}$
accurately.
Roles of $^6$Li breakup in the initial channel and
$^{17}$O breakup in the final channel are investigated with
the continuum-discretized coupled-channels
method (CDCC)~\cite{CDCC1,CDCC2}.
As shown in \S\ref{sec3-2}, the former is found important as a
large back-coupling to the elastic channel, while the latter is
confirmed much less important.
CDCC was proposed and developed by Kyushu group and has been
highly successful in quantitatively reproducing observables
of reaction processes in which virtual or real breakup effects of
the projectile are significant.~\cite{Ogata1,Ogata2}
CDCC treats continuum states of the
projectile nonperturbatively, with reasonable truncation and
discretization, and thus can describe the breakup effects
with very high accuracy. Note that theoretical foundation of CDCC
was established in Refs.~\citen{AYK,AKY,Piya}.
The transition from the $^{6}$Li$+^{13}$C channel to the $d+^{17}$O$^*$
channel is described with Born approximation; the breakup states of
$^6$Li are explicitly taken into account in the calculation of
the transfer process. The ANC thus extracted is compared with
the result of the previous DWBA analysis.

This paper is constructed as follows. In \S\ref{sec2} we formulate
the three-body wave functions in the initial
and final channels and the transfer cross section of the
$^{13}$C($^{6}$Li$,d$)$^{17}$O$^*$ reaction.
Numerical setting is described in \S\ref{sec3-1}.
Breakup effects of $^6$Li and $^{17}$O are investigated
in \S\ref{sec3-2}, and the transfer cross section is
analyzed and the ANC is extracted in \S\ref{sec3-3}.
In \S\ref{sec3-4} we see the convergence of the modelspace
of CDCC, and in \S\ref{sec3-5} we discuss the present result
in comparison with the previous DWBA result.
Finally, we give a summary in \S\ref{sec4}.

\section{Formulation}
\label{sec2}

%
%%%%%%%%%%%%%%%%%%%%%%%
%%%  Figure 1
%%%%%%%%%%%%%%%%%%%%%%%
\begin{figure}[htb]
\begin{center}
\includegraphics[width=0.9\textwidth,clip]{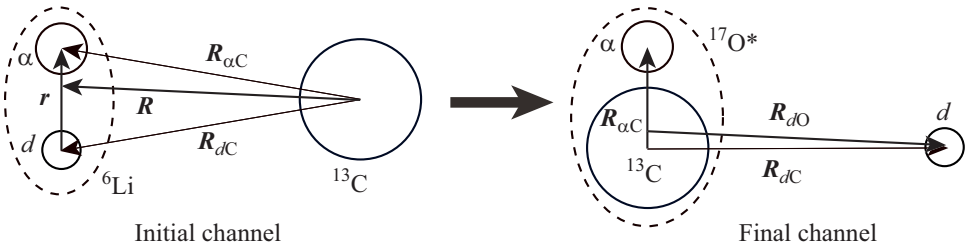}
\caption{Illustration of the three-body system in the
initial and final channels.}
\label{fig1}
\end{center}
\end{figure}
In the present calculation, we work with the $\alpha+d+^{13}$C model
shown in Fig.~\ref{fig1}.
The transition matrix ($T$ matrix) for the transfer reaction
$^{13}$C($^{6}$Li$,d$)$^{17}$O$^*$
is given by
\begin{equation}
T_{fi}=S_{\rm exp}^{1/2}
\left\langle
\Psi_f^{(-)}
\left|
V_{\rm tr}
\right|
\Psi_i^{(+)}
\right\rangle,
\label{tfi}
\end{equation}
where $\Psi_i^{(+)}$ and $\Psi_f^{(-)}$ are the three-body
wave functions of the system in the initial and final channels, respectively,
and $V_{\rm tr}$ is the transition operator of the transfer process.
We put a normalization constant $S_{\rm exp}^{1/2}$ in $T_{fi}$,
physics meaning of which is discussed below.

The three-body wave function $\Psi_i^{(+)}$ in the initial state
satisfies the Schr\"odinger equation
\beq
(H_i-E)\Psi_i^{(+)}({\bm r}, {\bm R})=0,
\label{sch1}
\eeq
where $E$ is the total energy of the system in the center-of-mass (c.m.)
frame and ${\bm r}$ (${\bm R}$) is the coordinate of $\alpha$
($^6$Li) relative to $d$ ($^{13}$C). The Hamiltonian $H_i$ is
given by
\beq
H_i=T_{\bm R}+V_{d {\rm C}}^{({\rm N})}(R_{d{\rm C}})
+V_{\alpha {\rm C}}^{({\rm N})}(R_{\alpha {\rm C}})
+V^{\rm Coul}(R)
+h_i,
\label{hi}
\eeq
where $T_{\bm R}$ is the kinetic energy operator associated with ${\bm R}$
and $h_i$ is the internal Hamiltonian of $^6$Li. We use
$V_{\rm XY}^{({\rm N})}$
for the nuclear interaction between X and Y; each of X and Y represents
a particle, i.e., $d$, $\alpha$, or C ($^{13}$C). Similarly,
${\bm R}_{\rm XY}$ denotes the relative coordinate between X and Y.
$V^{\rm Coul}$ is the Coulomb interaction between $^6$Li and $^{13}$C.
Note that we neglect the Coulomb breakup of $^6$Li, which can be
justified by the fact that the effective charge of the $\alpha+d$
system for electric dipole transition is almost zero.
Furthermore, as shown in \S\ref{sec3-2}, it is numerically confirmed that Coulomb breakup
processes due to electric quadrupole and higher multipoles are
negligibly small.

As the partial wave $\Psi_{i;JM}$ of $\Psi_i^{(+)}$, we adopt
the following CDCC wave function:
\beq
\Psi_{i;JM}^{\rm CDCC}({\bm r}, {\bm R})
=
\sum_{j=0}^{j_{\rm max}}
\sum_{\ell=0}^{\ell_{\rm max}}
\sum_{L=|J-\ell|}^{J+\ell}
\frac{\hat{\phi}_{j,\ell} (r)}{r}
\frac{\hat{\chi}_{j,\ell,L}^J (R)}{R}
\left[
i^\ell Y_\ell (\hat{\bm r})
\otimes
i^L Y_L (\hat{\bm R})
\right]_{JM},
\label{cdccwf}
\eeq
where $J$ and $M$ are the total angular momentum and its $z$-component,
respectively, and $\ell$ ($L$) is the orbital angular momentum between
$\alpha$ and $d$ ($^6$Li and $^{13}$C). We disregard the intrinsic spin
of each particle for simplicity.
The radial part of the $^6$Li wave function is denoted by
$\hat{\phi}_{j,\ell} (r)/r$, where $j$ is the energy index;
$j=0$ corresponds to the ground state and $j \neq 0$ to discretized
continuum states obtained by the momentum-bin discretization.~\cite{CDCC1}
The internal wave function $\hat{\Phi}_{j,\ell,m}$ given by
\beq
\hat{\Phi}_{j,\ell,m}({\bm r})
=
\frac{\hat{\phi}_{j,\ell} (r)}{r} i^\ell Y_{\ell m}(\hat{\bm r})
\eeq
satisfies
\beq
\left\langle
\hat{\Phi}_{j',\ell',m'}({\bm r})
\left|
h_i
\right|
\hat{\Phi}_{j,\ell,m}({\bm r})
\right\rangle
=
\epsilon_{j,\ell}\delta_{j'j}\delta_{\ell' \ell}\delta_{m'm}.
\label{onp}
\eeq

Inserting Eqs.~(\ref{hi}) and (\ref{cdccwf}) into Eq.~(\ref{sch1})
and making use of Eq.~(\ref{onp}), one obtains the following CDCC
equation:
\beq
\left[
-\frac{\hbar^2}{2\mu}\frac{d^2}{dR^2}
+\frac{\hbar^2}{2\mu}\frac{L(L+1)}{R^2}
+V^{\rm Coul}(R)
-E_{j,\ell}
\right]
\hat{\chi}_{c}^J (R)
=
-\sum_{cc'} F_{cc'}(R) \hat{\chi}_{c'}^J (R),
\label{cdcceq}
\eeq
where $\mu$ is the reduced mass of the $^6$Li-$^{13}$C system,
$E_{j,\ell}=E-\epsilon_{j,\ell}$, and
\beq
F_{cc'}(R)=
\left\langle
\frac{\hat{\phi}_{j',\ell'} (r)}{r}
\left[
i^\ell Y_\ell
\otimes
i^L Y_L
\right]_{JM}
\left|
V_{d {\rm C}}^{({\rm N})}
+V_{\alpha {\rm C}}^{({\rm N})}
\right|
\frac{\hat{\phi}_{j,\ell} (r)}{r}
\left[
i^{\ell'} Y_{\ell'}
\otimes
i^{L'} Y_{L'}
\right]_{JM}
\right\rangle_{{\bm r},\hat{\bm R}}.
\label{ff}
\eeq
For simple notation, we denote the channel indices $\{j,\ell,L \}$ as
$c$. The CDCC equation is solved numerically up to $R=R_{\rm max}$ and
$\hat{\chi}_{c}$ is connected with the usual boundary condition
\beq
\hat{\chi}_{c}^J(R)
\to
\left\{
\begin{array}
[c]{ll}%
U^{(-)}_{L,\eta_{j,\ell}} (K_{j,\ell} R)\delta_{cc_0}
-\sqrt{K_{0,\ell_0}/K_{j,\ell}}
\hat{S}_{cc_0}^J
U^{(+)}_{L,\eta_{j,\ell}} (K_{j,\ell} R)
& {\rm for } \; E_{j,\ell} \ge 0 \\
-\hat{S}_{cc_0}^J
W_{-\eta_{j,\ell},L+1/2} (-2i K_{j,\ell} R)
& {\rm for } \; E_{j,\ell} < 0
\end{array}
\right.,
\label{bc}
\eeq
where $K_{j,\ell}=\sqrt{2 \mu E_{j,\ell}}/\hbar$,
$U^{(-)}_{L,\eta_{j,\ell}}$ ($U^{(+)}_{L,\eta_{j,\ell}}$) is the
incoming (outgoing) Coulomb wave function with the Sommerfeld
parameter $\eta_{j,\ell}$, and
$W_{-\eta_{j,\ell},L+1/2}$ is the Whittaker function.
The subscript 0 of $\ell$ and $c$ represents the incident channel.
With the $S$-matrix elements $\hat{S}_{cc_0}^J$ in Eq.~(\ref{bc}),
one may obtain
any physics quantities with the standard procedure except that
one needs to make the discrete results smooth when breakup
observables are investigated.

Since the CDCC wave function $\Psi_i^{{\rm CDCC}}$
can be regarded as, with very high accuracy, an exact
solution to Eq.~(\ref{sch1}) in evaluation of $T$-matrix elements
that contain a short range interaction,
one may define $V_{\rm tr}$
by
\beq
V_{\rm tr}=V_{\alpha d}(r)
+V_{d {\rm C}}(R_{d{\rm C}})
+V_{\alpha {\rm C}}(R_{\alpha {\rm C}})
-V_{\rm aux}
\label{vres}
\eeq
with any choice of the auxiliary potential $V_{\rm aux}$.
In Eq.~(\ref{vres}), $V_{\alpha d}$,
$V_{d {\rm C}}$, and
$V_{\alpha {\rm C}}$ contain both
nuclear and Coulomb parts.
Note that $V_{\rm aux}$ determines the final state wave function
$\Psi_f^{(-)}$.
In the present study, we adopt
\beq
V_{\rm aux}=V_{d {\rm C}}(R_{d{\rm C}})
+V_{\alpha {\rm C}}(R_{\alpha {\rm C}})
+V_{\alpha d}^{({\rm C})}(r),
\label{vaux}
\eeq
which trivially gives
\beq
V_{\rm tr}=V_{\alpha d}^{({\rm N})}(r).
\eeq
The superscript (C) of $V_{\alpha d}$ in Eq.~(\ref{vaux})
represents the
Coulomb part of the interaction. We then have
\beq
(H_f-E)\Psi_f^{(+)}({\bm R}_{\alpha {\rm C}}, {\bm R}_{d {\rm O}})=0
\label{sch2}
\eeq
with
\beq
H_f=
T_{{\bm R}_{d {\rm O}}}
+V_{d {\rm C}}(R_{d{\rm C}})
+V_{\alpha d}^{({\rm C})}(r)
+h_f,
\label{hf}
\eeq
where $T_{{\bm R}_{d {\rm O}}}$ is the kinetic energy regarding
${\bm R}_{d {\rm O}}$ and $h_f$ is the internal Hamiltonian
of $^{17}$O. Note that we here consider a Schr\"odinger equation
for $\Psi_f^{(+)}$, which is the time-reversal of $\Psi_f^{(-)}$.

One can easily obtain the form of $\Psi_f^{(+)}$ based on CDCC,
$\Psi_f^{{\rm CDCC}(+)}$,
just in the same way as in the initial channel, except that i) we
should include Coulomb breakup of $^{17}$O, ii) we have no nuclear
part of $V_{\alpha d}$, and iii) the bound state of $^{17}$O
at 6.356~MeV is a p-wave that generates both
monopole and quadrupole interactions between $d$ and $^{17}$O;
the latter causes also change in the $d$-$^{17}$O angular
momentum that is called reorientation.
Note that $V_{d {\rm C}}$
in Eq.~(\ref{hf}) contains both nuclear and Coulomb parts, as
mentioned above.

It is shown in \S\ref{sec3-2} that $^{17}$O breakup channels have
very small ($\sim 5$\%) effects on the $d$-$^{17}$O elastic
scattering. Furthermore, the quadrupole interaction is found
negligibly small (see Fig.~\ref{fig2}).
Then we can approximate
\beq
\Psi_f^{{\rm CDCC}(-)}
\approx \varphi_0({\bm r})\xi_0^{(-)}({\bm R}_{d {\rm O}})
\equiv \Psi_f^{{\rm 1ch}(-)},
\label{final-wf}
\eeq
where $\varphi_0({\bm r})$ is the relative wave function between
$\alpha$ and $^{13}$C in $^{17}$O$^*$, and
$\xi_0^{(-)}({\bm R}_{d {\rm O}})$ is the distorted wave function obtained
by the single-channel calculation, in which
both the breakup channels and the aforementioned quadrupole interaction
are switched off.

In the calculation of $T_{fi}$, we make zero-range approximation;
the strength $D_{j,\ell}$ of the zero-range $\alpha$-$d$ interaction
is given by
\beq
D_{j,\ell}=
\int
\hat{\phi}^*_{j,\ell} (r)
V_{\alpha d}^{({\rm N})}(r)
\hat{\phi}_{j,\ell} (r)
dr.
\eeq
The finite-range correction to the zero-range calculation of
$T_{fi}$ is made with the standard prescription.~\cite{Satchler}
One may examine the validity of this approximation by the magnitude
of the correction. We use $\Psi_i^{(+)}$ calculated with CDCC,
while $\Psi_f^{{\rm 1ch}(-)}$ of Eq.~(\ref{final-wf}) is adopted
as $\Psi_f^{(-)}$, in the evaluation of $T_{fi}$.

\section{Results and discussion}
\label{sec3}

\subsection{Numerical input}
\label{sec3-1}

The $\alpha$-$d$ wave function in $\Psi_{i}^{\rm CDCC}$
is constructed by following Ref.~\citen{Sakuragi}, except that
we do not use the orthogonal condition model (OCM) but
exclude Pauli's forbidden states by hand.
We include $\ell=0$, 1, and 2 states.
As for the nuclear part of the $\alpha$-$d$ interaction for $\ell=0$,
we use
\beq
V_{\alpha d; \ell=0}^{({\rm N})}(r)
=
-105.5 \exp[-(r/2.191)^2]
+46.22 \exp[-(r/1.607)^2].
\eeq
For $\ell=2$,
\beq
V_{\alpha d; \ell=2}^{({\rm N})}(r)
=
-85.00 \exp[-(r/2.377)^2]
+30.00 \exp[-(r/1.852)^2]
\eeq
is adopted. We neglect the intrinsic spin $S$ of $d$, and we have only
one resonance state at 3.474~MeV (measured from the ground state energy)
with a width of 0.45~MeV. It is
found that, however, if we include $S$ and a spin-orbit interaction
that reproduces the $1^+$, $2^+$, and $3^+$ resonance states,
the resulting value of the ANC shown below changes by only 0.2\%.
Thus, the separation of the $\ell=2$ resonance state
to $1^+$, $2^+$, and $3^+$ resonance states
by the spin-orbit interaction plays no role
in the present subbarrier $\alpha$ transfer reaction.
For $\ell=1$, we adopt~\cite{Matsumoto}
\beq
V_{\alpha d}^{({\rm N})}(r)
=-74.19 \exp[-(r/2.236)^2],
\eeq
which is used also for $\ell>2$ when we check the convergence
of CDCC calculation with respect to $\ell_{\rm max}$ (see \S\ref{sec3-4}).
The Coulomb interaction between $\alpha$ and $d$ is evaluated by
assuming a uniformly charged sphere with the charge radius
$R_{\mathrm{C}}$ of 3.0~fm; see Eq.~(\ref{vcoul}) below.

We take the maximum value $k_{\rm max}$ ($r_{\rm max}$) of the relative
wave number $k$ (coordinate $r$) between $\alpha$ and $d$ to be
2.0~fm$^{-1}$ (60~fm); the maximum relative energy $\epsilon_{\rm max}$ is
62.4~MeV.
We use $j_{\rm max}=100$ for each of the
$\ell=0$, 1, and 2 states and the width $\Delta k$ of the momentum bin
is thus 0.02~fm$^{-1}$. The number of channels, $N_{\rm ch}$, in the
CDCC equation (\ref{cdcceq}) is 601.
When we see the effects of Coulomb breakup in Fig.~\ref{fig2},
we take $r_{\rm max}=300$~fm.

As for the interactions of the $\alpha$-$^{13}$C and
$d$-$^{13}$C systems, we use the parameters shown in Table~\ref{tab1}.
The standard
Woods-Saxon form is adopted:
\beq
V(x)=-V_0 f_{\rm V} (x)
-iW_0 f_{\rm W} (x) + V_{\rm C}(x),
\eeq
where
$f_{\rm V} (x)=(1+\exp[(x-R_{\rm V})/a_{\rm V}])^{-1}$
and
$f_{\rm W} (x)=(1+\exp[(x-R_{\rm W})/a_{\rm W}])^{-1}$.
The Coulomb interaction $V_{\rm C}(x)$ is given by
\beq
V_{\rm C}(x)=
\left\{
\begin{array}
[c]{ll}%
\displaystyle\frac {Z_1 Z_2e^{2}}{2R_{\mathrm{C}}}\left(
3-\frac{x^{2}}{R_{\mathrm{C}}^{2}}\right)  & \quad x\leq R_{\mathrm{C}}\\
\displaystyle\frac{Z_1 Z_2e^{2}}{x} & \quad
x>R_{\mathrm{C}}%
\end{array}
\right.,
\label{vcoul}
\eeq
where $Z_1 Z_2$ is the product of the atomic numbers of the
interacting particles. These parameters are used in the calculation
of both initial and final state wave functions. The parameter set for
the $d$-$^{13}$C system is determined to reproduce the elastic
scattering cross section obtained with the parameters
in Ref.~\citen{Eric1} that contains a spin-orbit part.
We determine $V_0$ for the $\alpha$-$^{13}$C system to reproduce
$\varepsilon_0$ assuming that
the orbital angular momentum is 1 and the number of forbidden states is 2.
Note that we use Eq.~(\ref{vcoul}) with $R_{\rm C}=2.94$~fm for
the $^6$Li-$^{13}$C Coulomb interaction unless we include
Coulomb breakup of $^6$Li.

%%%%%%%%%%%%%%%%%%%%%%%
%%%  Table I
%%%%%%%%%%%%%%%%%%%%%%%
\begin{table}[htb]
\caption{
Potential parameters used in the present calculation.
}
\begin{center}
\begin{tabular}{cccccccc}
\hline
\hline
System &
$V_0$ & $R_{\rm V}$ & $a_{\rm V}$ &
$W_0$ & $R_{\rm W}$ & $a_{\rm W}$ &
$R_{\rm C}$
\\
 &
(MeV) & (fm) & (fm) &
(MeV) & (fm) & (fm) &
(fm)
\\
\hline
$\alpha$+$^{13}$C &
69.30 & 2.939 & 0.670 &
---   & ---   & ---  &
2.969
\\
$d$+$^{13}$C &
73.05 & 3.128 & 0.780 &
10.50 & 2.986 & 0.800 &
2.969 \\
\hline
\hline
\end{tabular}
\label{tab1}
\end{center}
\end{table}

In the calculation of $\Psi_{i;JM}^{\rm CDCC}$, we use
$R_{\rm max}=15$~fm and $J_{\rm max}=7$.
Note that we explicitly include closed channels, in which
$E_{j,\ell}<0$, in CDCC calculations.
In the evaluation of $T_{fi}$,
we set the maximum value of $R_{d {\rm C}}$ to be 30~fm;
we use the asymptotic form of $\hat{\chi}_{c}^J$, Eq.~(\ref{bc}),
to obtain $\Psi_{i;JM}^{\rm CDCC}$ for $R > 15$~fm.
When we include Coulomb breakup, we set $R_{\rm max}$ to 200~fm.

For the final channel, the relative energy between $\alpha$ and
$^{13}$C in the $1/2^+$ state at 6.356~MeV is
$\varepsilon_0=-3$~keV from the $\alpha$-$^{13}$C threshold.
In the calculation of $\Psi_f^{{\rm CDCC}(-)}$,
we include the p-wave bound state and the s-, p-, d-continua of
the $\alpha$+$^{13}$C system up to the relative momentum
of 1.2~fm$^{-1}$ (relative energy of 39.6~MeV)
with the momentum bin with a common width 0.06~fm$^{-1}$.
The maximum values of $R_{\alpha {\rm C}}$ and $R_{d {\rm O}}$
are both set to 100~fm, and we put $J_{\rm max}=10$.
We include all closed channels in the CDCC calculations as in
the initial channel.

\subsection{Breakup effects of $^6$Li and $^{17}$O}
\label{sec3-2}

%
%%%%%%%%%%%%%%%%%%%%%%%
%%%  Figure 2
%%%%%%%%%%%%%%%%%%%%%%%
\begin{figure}[htbp]
%\begin{figure}[t]
\begin{center}
 \includegraphics[width=1.0\textwidth,clip]{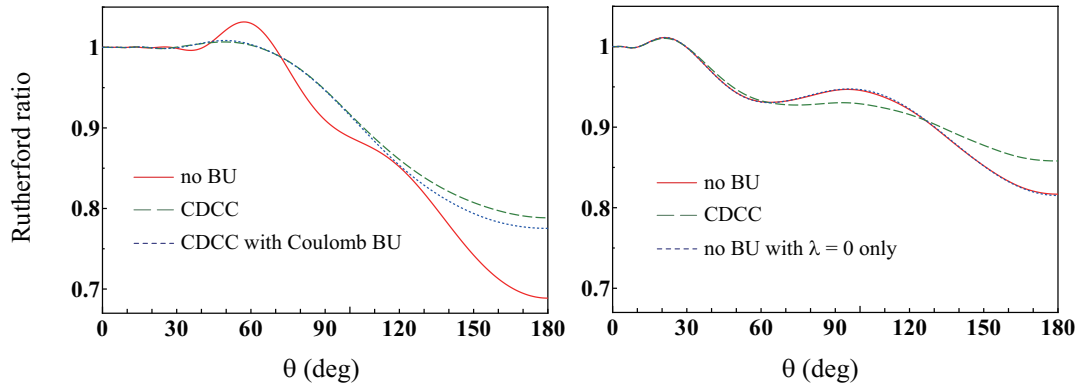}
 \caption{
 (color online)
Elastic cross sections of
$^6$Li-$^{13}$C at 3.6~MeV (left panel) and $d$-$^{17}$O$^*$
at 1.1~MeV (right panel). In each panel, the dashed line shows
the result of CDCC and the solid line is the result without
breakup channels. The dotted line in the left panel is the result
of CDCC with both nuclear and Coulomb breakup, while that in the
right panel shows the result without breakup including only the
monopole interaction between $d$ and $^{17}$O.
}
\label{fig2}
\end{center}
\end{figure}
Figure~\ref{fig2} shows the elastic cross sections of
$^6$Li-$^{13}$C at 3.6~MeV (left panel) and $d$-$^{17}$O$^*$
at 1.1~MeV (right panel) corresponding to the initial and
final channels, respectively, of the
$^{13}$C($^{6}$Li$,d$)$^{17}$O$^*$ reaction.
In each panel, the dashed line shows
the result of CDCC and the solid line is the result without
breakup channels.
One sees from the left panel significant breakup effects on
the elastic cross section, i.e., a large back-coupling to the
elastic channel. Another finding is the inclusion of Coulomb
breakup (the dotted line in the left panel)
little affects the cross section. One can thus infer
that nuclear breakup plays important roles in the
$^{13}$C($^{6}$Li$,d$)$^{17}$O$^*$ reaction, and conclude that
neglect of
the Coulomb breakup in the calculation of $\Psi_{i;JM}^{\rm CDCC}$
is justified. On the other hand, in the final channel,
effects of nuclear and Coulomb breakup are found very small as shown
in the right panel;
they change the cross section for $\theta \ga 60^\circ$
by 5\% at most. We further investigate the breakup effects on
the $d$-$^{17}$O$^*$ wave function in the elastic channel.
The absolute value (argument) of the wave function for $J=0$
at $R_{d {\rm O}}=10$~fm, which is found to have the main
contribution to the transfer amplitude, is 0.982 and 0.956
($278.8^\circ$ and $276.7^\circ$) when the breakup states
of $^{17}$O are included and neglected, respectively;
the breakup effects are about 3\%.
Therefore, we can disregard the breakup channels
of the $d$-$^{17}$O system in the calculation of $T_{fi}$
with the error of 5\% at most. The dotted line in the right panel
shows the result with neglecting both the breakup channels and
the quadrupole interaction between $d$ and $^{17}$O,
which is almost identical to the solid line. Thus, one
can use Eq.~(\ref{final-wf}) in the calculation of the
final state wave function; we estimate the error due to
this approximation to be 5\% as mentioned above.
It should be noted that breakup cross sections in the
initial and final channels are both found smaller than
the nuclear part of the elastic cross section by
about four orders of magnitude.

The very small breakup effects in the final channel are
because the incoming energy of $d$ is suitably
below the Coulomb barrier, and the interaction that causes breakup in
Eq.~(\ref{hf}) is significantly weaker than in Eq.~(\ref{hi}); note that
$V_{\alpha d}^{({\rm N})}(r)$ is defined as $V_{\rm tr}$ and does not
appear in Eq.~(\ref{hf}).

\subsection{Transfer cross section and ANC}
\label{sec3-3}

%
%%%%%%%%%%%%%%%%%%%%%%%
%%%  Figure 3
%%%%%%%%%%%%%%%%%%%%%%%
\begin{figure}[htbp]
%\begin{figure}[t]
\begin{center}
 \includegraphics[width=0.6\textwidth,clip]{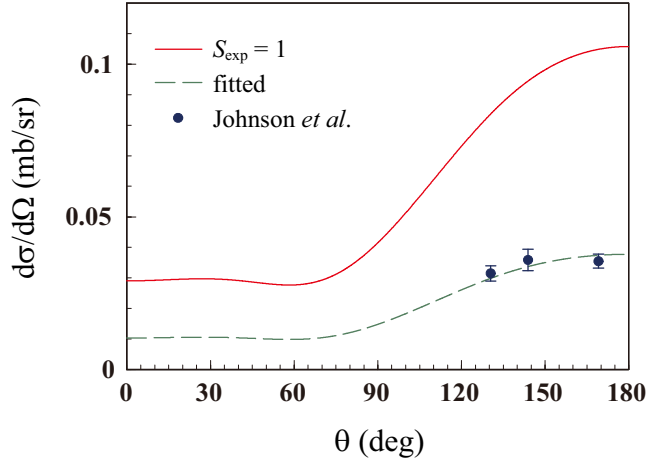}
 \caption{
 (color online)
Cross section of the transfer reaction
$^{13}$C($^{6}$Li$,d$)$^{17}$O$^*$
at 3.6~MeV. The solid line is the result of
calculation with $S_{\rm exp}=1$. The dashed line
is the result of the $\chi^2$ fit to the experimental
data taken from Ref.~\citen{Eric1}.
 }
\label{fig3}
\end{center}
\end{figure}
We show in Fig.~\ref{fig3} the cross section of the transfer reaction
$^{13}$C($^{6}$Li$,d$)$^{17}$O$^*$ at 3.6 MeV as a function
of the outgoing angle $\theta$ of $d$ in the c.m. frame.
The solid line represents the result with $S_{\rm exp}=1$
and the dashed line shows the result of the $\chi^2$ fit
to the experimental data.~\cite{Eric1}
The resulting value of $S_{\rm exp}$ is 0.357.
Note that $S_{\rm exp}$
cannot be regarded as a spectroscopic factor. Indeed, $S_{\rm exp}$ has
strong dependence on the model wave function
of the $\alpha$-$^{13}$C system; typically, it varies by a factor of 2
with changing the geometric parameters
of $V_{\alpha {\rm C}}^{({\rm N})}$ by 30\%.
This clearly shows that it is not feasible to determine $S_{\rm exp}$
from the present analysis of the experimental data.
On the other hand, the ANC
$C_{\alpha ^{13}\mbox{\scriptsize C}}^{^{17}\mbox{\scriptsize O}^*}$
given by
\beq
C_{\alpha ^{13}\mbox{\scriptsize C}}^{^{17}\mbox{\scriptsize O}^*}
=
S_{\rm exp}^{1/2}\,
C_{\alpha ^{13}\mbox{\scriptsize C}}^{\rm sp}
\eeq
with the single particle ANC
$C_{\alpha ^{13}\mbox{\scriptsize C}}^{\rm sp}$
of the $\alpha$-$^{13}$C wave function, is robust against
changes in the potential parameters.
This shows that the reaction process considered is peripheral
with respect to $R_{\alpha {\rm C}}$, i.e., only the tail of the
$\alpha$-$^{13}$C wave function contributes to the transition amplitude.
Note that $C_{\alpha ^{13}\mbox{\scriptsize C}}^{\rm sp}$
is defined by
\begin{equation}
C_{\alpha ^{13}\mbox{\scriptsize C}}^{\rm sp}
=
\dfrac{
R_{\alpha {\rm C}}\,{\bar{\varphi}_0(R_{\alpha {\rm C}})}
}
{
W_{-\bar{\eta},3/2}(2 \kappa_0 R_{\alpha {\rm C}})
\Gamma(2+\bar{\eta})
}
\;\; {\rm at} \;\; R_{\alpha {\rm C}} \gg R_{\rm N},
\end{equation}
where $\bar{\varphi}_0$ is the radial part of
$\varphi_0$, $\bar{\eta}$ is the Sommerfeld parameter of the
$\alpha$-$^{13}$C system,
$\kappa_0=\sqrt{-2\mu_{\alpha ^{13}\mbox{\scriptsize C}}\,\varepsilon_0}/\hbar$
with $\mu_{\alpha ^{13}\mbox{\scriptsize C}}$ the reduced mass of
$\alpha$ and $^{13}$C, $\Gamma$ is the Gamma function, and
$R_{\rm N}$ represents the range of $V_{\alpha {\rm C}}^{({\rm N})}$.

The value of
$(C_{\alpha ^{13}\mbox{\scriptsize C}}^{^{17}\mbox{\scriptsize O}^*})^2$
extracted by the present calculation is
1.03~fm$^{-1}$.
We then evaluate the uncertainty of
$(C_{\alpha ^{13}\mbox{\scriptsize C}}^{^{17}\mbox{\scriptsize O}^*})^2$
associated with the $\alpha$-$^{13}$C and $d$-$^{13}$C
potential parameters shown in Table~\ref{tab1}
by changing each value by 30\%. Note that $V_0$ for
$\alpha$-$^{13}$C has a constraint that it must reproduce
$\varepsilon_0$.
The uncertainty is found to be 22\%.
We take into account also the uncertainty  due to the use
of Eq.~(\ref{final-wf}) (5\%) and that coming from the zero-range
approximation to $V_{\alpha d}^{({\rm N})}$ (8\%), and conclude
that the theoretical uncertainty is totally 24\%.
Including the ambiguity of experimental information~\cite{Eric1}
together, we finally obtain
\beq
(C_{\alpha ^{13}\mbox{\scriptsize C}}^{^{17}\mbox{\scriptsize O}^*})^2
=1.03 \pm 0.25 \; {\rm (theor)}\; \pm 0.15 \; {\rm (expt)},
\label{ANC-result}
\eeq
where (theor) and (expt) respectively represent theoretical and
experimental uncertainties.

\subsection{Convergence of the CDCC wave function in the initial channel}
\label{sec3-4}

%
%%%%%%%%%%%%%%%%%%%%%%%
%%%  Figure 4
%%%%%%%%%%%%%%%%%%%%%%%
\begin{figure}[htbp]
%\begin{figure}[t]
\begin{center}
 \includegraphics[width=1.0\textwidth,clip]{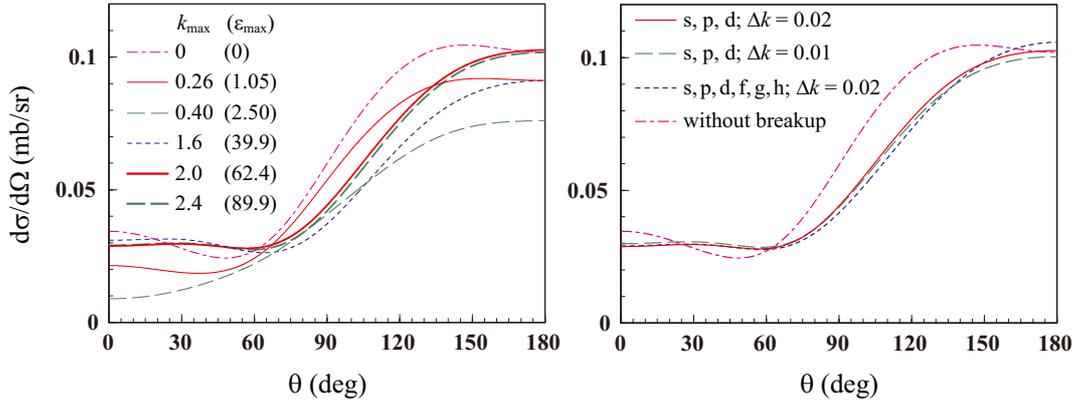}
 \caption{
 (color online)
The dependence the cross section for
$^{13}$C($^{6}$Li$,d$)$^{17}$O$^*$ at 3.6~MeV
on the modelspace of CDCC for $\Psi_i^{{\rm CDCC}}$.
In the left panel, $k_{\rm max}$ is varied.
The values of $k_{\rm max}$ are shown in unit of fm$^{-1}$ and
the corresponding values of $\epsilon_{\rm max}$
are given in the parentheses in unit of MeV.
Here, the $\ell=0$, 1, and 2 breakup continua are taken with
$\Delta k=0.02$~fm$^{-1}$.
In the right panel, the dashed line stands for the result
of the $\ell=0$, 1, and 2 breakup continua
with $k_{\rm max}=2.0$~fm$^{-1}$ and $\Delta k=0.01$~fm$^{-1}$.
The dotted line shows the
result of the $0 \le \ell \le 5$ continua
with $k_{\rm max}=2.0$~fm$^{-1}$ and $\Delta k=0.02$~fm$^{-1}$.
The thick solid line (the solid line) in the left (right) panel
is the result of
the $\ell=0$, 1, and 2 breakup continua
with $k_{\rm max}=2.0$~fm$^{-1}$ and $\Delta k=0.02$~fm$^{-1}$
and the same as the solid line in Fig.~\ref{fig3}.
The result without breakup channels is also shown by the dash-dotted
line in each panel.
}
\label{fig4}
\end{center}
\end{figure}

Figure~\ref{fig4} shows the dependence of
the cross section for
$^{13}$C($^{6}$Li$,d$)$^{17}$O$^*$ at 3.6~MeV
on the modelspace of CDCC for $\Psi_i^{{\rm CDCC}}$.
In the left panel, we show the convergence of the cross section
with respect to increasing $k_{\rm max}$, where
the $\ell=0$, 1, and 2 breakup continua are taken with
$\Delta k=0.02$~fm$^{-1}$.
One can see that the convergence is very slow and obtained
at $k_{\rm max}=2.0$~fm$^{-1}$.
In usual CDCC calculation,
one takes only the open channels, i.e., channels with $E_{j,\ell}>0$.
The result thus obtained (the thin solid line) is, however,
sizably different from  the converged one (the thick dotted line),
at backward angles in particular.
Thus, inclusion of the breakup channels is important.

In the right panel of Fig.~\ref{fig4}, the dashed line is the result
including the $\ell=0$, 1, and 2 breakup continua with
$\Delta_k=0.01$~fm$^{-1}$ and $k_{\rm max}=2.0$~fm$^{-1}$
($N_{\rm ch}=1201$), and the dotted line is the result including
the $\ell=0,$ 1, 2, 3, 4, and 5 breakup continua
with $\Delta_k=0.02$~fm$^{-1}$ and $k_{\rm max}=2.0$~fm$^{-1}$
($N_{\rm ch}=2101$).
The dashed and solid lines both agree well with the solid line,
which is the same as in Fig.~\ref{fig3}. In fact, the resulting values
of $(C_{\alpha ^{13}\mbox{\scriptsize C}}^{^{17}\mbox{\scriptsize O}^*})^2$
differ from each other by less than 1\%.
Thus, the modelspace used in the solid line of Fig.~\ref{fig3}
gives good
convergence of the calculated cross section, hence
$C_{\alpha ^{13}\mbox{\scriptsize C}}^{^{17}\mbox{\scriptsize O}^*}$.

\subsection{Discussion on the comparison with the previous DWBA analysis}
\label{sec3-5}

%
%%%%%%%%%%%%%%%%%%%%%%%
%%%  Figure 5
%%%%%%%%%%%%%%%%%%%%%%%
\begin{figure}[htbp]
%\begin{figure}[t]
\begin{center}
 \includegraphics[width=0.6\textwidth,clip]{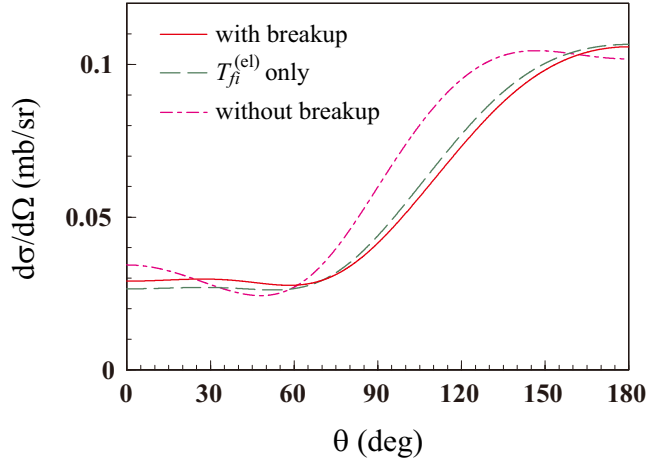}
 \caption{
 (color online) $^6$Li-breakup effect on
the transfer reaction
$^{13}$C($^{6}$Li$,d$)$^{17}$O$^*$
at 3.6~MeV.
The solid and dash-dotted lines show the results
of calculations with and without $^6$Li breakup channels,
respectively.
The dashed line stands for the result of the elastic transfer process only.
The solid and dash-dotted lines are respectively the same as those
in the right panel of Fig.~\ref{fig4}.
}
\label{fig5}
\end{center}
\end{figure}

Figure~\ref{fig5} shows the $^6$Li-breakup effect on the
transfer reaction.
The solid and dash-dotted lines are the results of
the calculation with and without the breakup channels, respectively;
they are shown also in the right panel of Fig.~\ref{fig4}.
The two results largely deviate from each other, indicating that
the breakup effect is important, as mentioned in \S\ref{sec3-4}.
This does not necessarily mean, however, inadequacy of DWBA, as discussed below.
The transition matrix elements of the transfer reaction can be separated
into two parts,
\bea
T_{fi}=S_{\rm exp}^{1/2}
\left[
T_{fi}^{\rm (el)}+T_{fi}^{\rm (br)}
\right]
\label{tfi-separation-1}
\eea
with
\bea
T_{fi}^{\rm (el)}&=&
\left\langle
\Psi_f^{(-)}
\left|
V_{\rm tr}
\right|
\Psi_{\rm el}^{(+)}
\right\rangle ,
\\
T_{fi}^{\rm (br)}&=&
\left\langle
\Psi_f^{(-)}
\left|
V_{\rm tr}
\right|
\Psi_{\rm br}^{(+)}
\right\rangle ,
\label{tfi-separation-2}
\eea
where $\Psi_{\rm el}^{(+)}$ and $\Psi_{\rm br}^{(+)}$ are the
elastic and breakup parts of the CDCC wave function $\Psi_i^{{\rm CDCC}}$.
The transition matrix $T_{fi}^{\rm (el)}$ describes the transfer reaction
from the elastic channel, i.e., the elastic transfer process,
which includes the back-coupling effect
of the breakup channels to the elastic channel.
On the other hand, $T_{fi}^{\rm (br)}$ describes the transfer reaction
from $^6$Li breakup channels, i.e., the breakup transfer process.
Thus, there are
two kinds of breakup effects on the transfer reaction;
one is the back-coupling effect in the elastic transfer process and
the other is the presence of the breakup transfer process.
The dashed line is a result of the elastic transfer transition only.
The result agrees with the solid line, indicating that
the breakup transfer transition is much smaller than the elastic transfer one.
This is consistent with the small breakup cross section
of $^6$Li by $^{13}$C as mentioned in \S\ref{sec3-2}.
Hence, only the back-coupling effect is important
in the present subbarrier transfer reaction.
In DWBA, the back-coupling effect is expected to be included
by using the $^6$Li optical potential, which
describes the elastic scattering by definition, as the distorting potential.
The ANC,
$(C_{\alpha ^{13}\mbox{\scriptsize C}}^{^{17}\mbox{\scriptsize O}^*})^2$,
 extracted in the preceding
DWBA calculation~\cite{Eric1} is $0.89 \pm 0.23$~fm$^{-1}$.
This value agrees well with the present result Eq.~(\ref{ANC-result})
within the uncertainties.

\section{Summary}
\label{sec4}

In summary, we analyze the
$^{13}$C($^{6}$Li$,d$)$^{17}$O(6.356~MeV, $1/2^+$)
reaction at 3.6~MeV by the three-body ($\alpha+d+^{13}$C) model.
The breakup effects of $^6$Li and $^{17}$O are investigated by
CDCC. Those of $^6$Li are found important as a large back-coupling
to the elastic channel, while those of $^{17}$O turns out
negligible with an error of 5\%.
The transfer cross section is calculated with Born approximation
to the transition interaction, and including only the breakup
of $^6$Li.
The ANC extracted by the three-body reaction model is
$
(C_{\alpha ^{13}\mbox{\scriptsize C}}^{^{17}\mbox{\scriptsize O}^*})^2
=
1.03 \pm 0.25 \; {\rm (theor)}\; \pm 0.15 \; {\rm (expt)}
$.
The back-coupling effect of $^{6}$Li breakup on the transfer reaction is
large, while the breakup transfer transition is negligible compared
with the elastic transfer transition.
The preceding DWBA calculation implicitly treated the back-coupling
effect by using a $^{6}$Li optical potential that described the elastic
scattering as the distorting  potential.
The value of
$(C_{\alpha ^{13}\mbox{\scriptsize C}}^{^{17}\mbox{\scriptsize O}^*})^2$
extracted by DWBA is $0.89 \pm 0.23$~fm$^{-1}$, which is
consistent with the present value within the uncertainties.
It can be conjectured that in the DWBA calculation, the aforementioned
back-coupling effect in the initial channel was properly
included.
However, this will not always be the case, since the optical
potential is determined phenomenologically. Furthermore, breakup
transfer processes may be important in other subbarrier $\alpha$
transfer reactions.
The present three-body approach, therefore, should
be applied systematically to these reactions.
From theoretical point of view, inclusion of CDCC wave functions
in both initial and final channels will be an important subject;
to achieve this, one should treat, in principle, very large coordinate
space in the calculation of the $T$ matrix, since there is no damping
in the overlap kernel.
It will be interesting to use four-body CDCC~\cite{Matsumoto2} based
on a $p+n+\alpha+^{13}$C model to obtain the wave function in the
initial channel. At this stage, however, the modelspace required is
too large for four-body CDCC to be applied.

\vspace{3mm}

One of the authors (K.~O.) wishes to thank G.~V.~Rogachev and E.~D.~Johnson
for valuable discussions and providing detailed information
on their DWBA calculation.
The authors are grateful to Y.~Iseri for providing a computer
code {\sc rana} for calculation of transfer processes.
The computation was carried out using the computer facilities at
the Research Institute for Information Technology, Kyushu University.

%%--------------------------------------------------------------------%%
%%                           References                               %%
%%--------------------------------------------------------------------%%

\end{document}